\documentclass[12pt]{article}
\usepackage{epsfig}

\parskip=\medskipamount 
plus.3em minus.5em \abovedisplayshortskip=.5em plus.2em minus.4em
\renewcommand{\thesection}{\arabic{section}}

\catcode`@=11

\@addtoreset{equation}{section} \@addtoreset{equation}{subsection}
\def\theequation{\ifnum\value{section}=0 \arabic{equation}\ignorespaces
\else \ifnum\value{section}=-1 A.\arabic{equation}\ignorespaces
\else \ifnum\value{subsection}=0
\thesection.\arabic{equation}\ignorespaces \else
\thesection.\arabic{subsection}.\arabic{equation}\ignorespaces
                             \fi
                        \fi
                   \fi}

{\catcode`\'=\active \def'{{}^\bgroup\prim@s}}

\catcode`@=12

\newcommand{\bq}{\begin{equation}}
\newcommand{\be}{\begin{equation}}
\newcommand{\fq}{\end{equation}}
\newcommand{\ee}{\end{equation}}
\newcommand{\bqr}{\begin{eqnarray}}
\newcommand{\beqs}{\begin{eqnarray}}
\newcommand{\fqr}{\end{eqnarray}}
\newcommand{\eeqs}{\end{eqnarray}}

\def\bop#1{\setbox0=\hbox{$#1M$}\mkern1.5mu
    \vbox{\hrule height0pt depth.04\ht0
    \hbox{\vrule width.04\ht0 height.9\ht0 \kern.9\ht0
    \vrule width.04\ht0}\hrule height.04\ht0}\mkern1.5mu}

\begin{document}
\thispagestyle{empty}

\vskip .6in
\begin{center}

{\bf Light Computing}

\vskip .6in

{\bf Gordon Chalmers}
\\[5mm]

{e-mail: gordon@quartz.shango.com}

\vskip .5in minus .2in

{\bf Abstract}

\end{center}

A configuration of light pulses is generated, together with emitters 
and receptors, that allows computing.  The computing is extraordinarily 
high in number of flops per second, exceeding the capability of a quantum 
computer for a given size and coherence region.  The emitters and 
receptors are based on the quantum diode, which can emit and detect 
individual photons with high accuracy.

\newpage

Traditional computing relies on the movement of electrons through logic 
circuits.  Planned designs for quantum computing \cite{QuantumComputing1} 
require the manipulation 
of wavefunctions through nanocircuitry, and this requires special equipment 
to manafacture.  These two types of circuitries are commonly described in the 
literature, with emphasis on the latter in its computational power.  There is 
another possibility using an aspect of quantum computing, that is coherence 
of electrons or photons with eachother or themselves.  Laser computing 
requires the coherence of many photons in a manner that short pulses in 
the overall wavefunction can accomplish the equivalent of a circuit.  
The concept and several basic architectures are presented.  

Laser computing does not require any physical circuits, except the emitters 
and receptors that are used in the beam configuration.  For example, 
consider a 
platform half a meter in diameter containing, possibly quantum, diodes 
emitting a cohered photon shower every picosecond.  There would be an 
approximate $10^{18}$ elements or less in the emitter.  Give these diodes an 
angular precision of $\Delta\theta=10^{-4}-10^{-6}$ or more.  The second 
platform is the receptor platform located a distance $d$ meters away.  
The receptor dish should be able to measure 
the amplitude and phase of the incoming cohered beam; if the 
frequencies are required to be measured then their components have to 
be separated and directed to the appropriate region in the receptor 
platform.  Either that or the receptor elements must distinguish 
the frequency of the incoming photon.  The parameters of the 
beam, such as the radius and the distance, should be chosen so that the entire 
beam has coherence with itself over the distance $d$.  

The type of quantum diodes that are used in the models are defined as a 
lattice of a material that has a series of donor atoms in a row 
which are meta-stable 
and form a current when a small current is applied.  The row of 
donated electrons are trapped in 
a meta-stable quantum well near the boundary of the material for a small 
period of time, where they coher and then escape, emitting a number 
of cohered photons.  The width and depth of the quantum well (trap) dictate 
the properties of the emitted burst of photons.  

\begin{figure}
\begin{center}
\epsfxsize=12cm
\epsfysize=6cm
\epsfbox{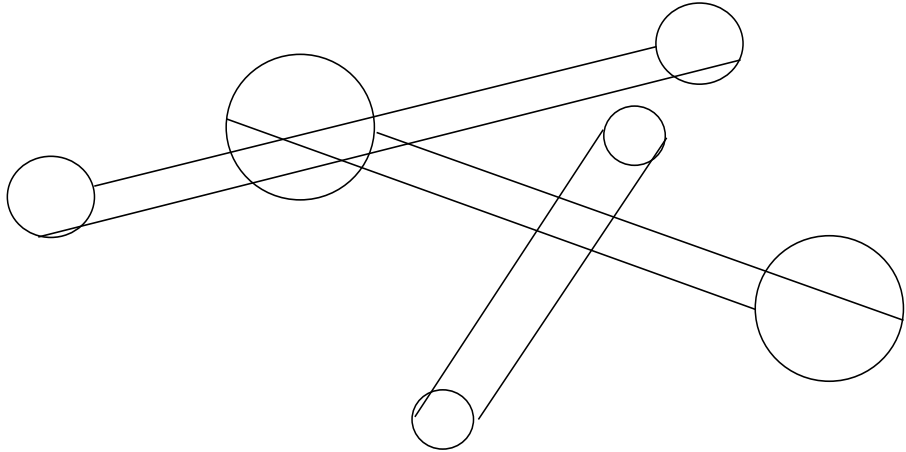}
\end{center}
\caption{A linear laser design.}
\end{figure}

Next, cohered beams are placed that crossfire the previous one; this is 
described in figure 1.  The angles and orientation are chosen so as to 
alter the coherence with eachother and the 'primary' beam.  Firing photons 
in patterns, and possibly with differing frequencies, from the various 
emitters in the emitter platform generate very complex coherence 
patterns in the primary beam and the secondary beams.  Consider ten beams 
a half meter long with a primary beam three meters.  The optimum case 
should not coher over the maximum and this puts a constraint 
on the geometry; the considered example is illustrative.  The secondary 
beams can alter the primary beam with an approximate $10^8$ (from the 
diameter) and maybe $10^4$ during a single cycle of the primary beam's 
traversing the distance $d$ (if the diode has picosecond timing).  Then 
there are ten beams, and the individual 
photons from the secondary beam quantum cohers with the entire primary beam.  
The complexity is overwhelming, and possibly equals or exceeds the concept 
of a quantum computer.  Considering the length $d=3$ meters 
the complexity with one secondary beam could be $10^{10^{12}}$, not 
including self-coherence of the two beams.  This grows 
when the coherence of the secondary beams with themselves and with the 
other beams are considered.  

The coherence length of the lasers should encompass the entire geometry for 
maximal effect.  If the former is smaller than the size of the system, 
degradation of the photon coherence occurs.  

A different configuration consists of a primary beam along an axis of a 
sphere.  Diodes are placed for maximal packing along the inner wall 
of the sphere.  This might lead to maximal coherence of the beam with the 
sphere's diodes.  Firing the diodes in various patterns would coher the 
beam in various ways, generating a very fast quantum computer.  

\begin{figure}
\begin{center}
\epsfxsize=10cm
\epsfysize=8cm
\epsfbox{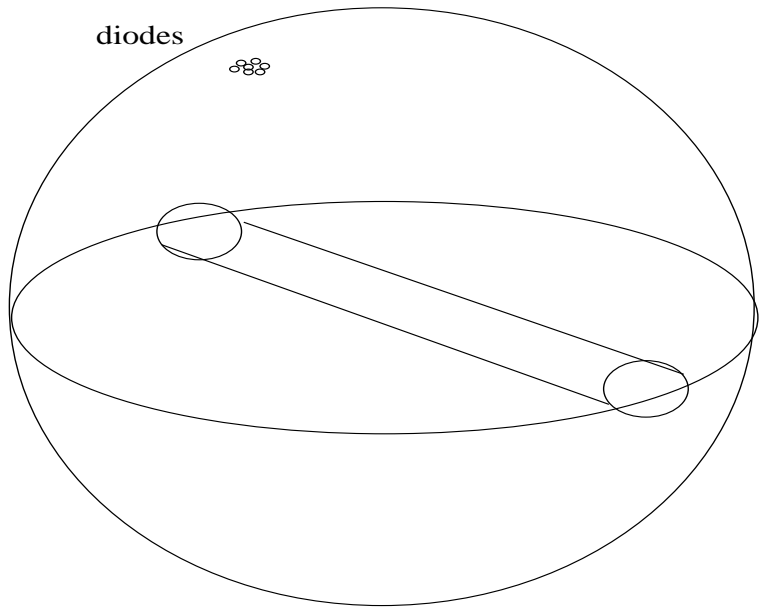}
\end{center}
\caption{A spherical arrangement.  The litte circles are the diode 
emitters and the tube is the primary beam.}
\end{figure}

A small set of beams with receptors and emitters can be used also in an 
arrangement of a lattice, such as a bipartite diamond configuration, that 
fits in a box on your desk, the size of a computer.  It would be encased 
in a material so that coherence with the outside is minimized.  The 
emitters and receptors span a smaller area.  A secondary fast chip can 
be used to pattern the emitters.  

The configuration of laser beams requires much accuracy and stability to 
ensure its quantum coherence as designed.  A question is whether the laser 
computing is more stable than a nano-sized circuit that captures the 
wavefunction of an electron in a complicated environment.   

One difficulty in programming the cohered laser computing involves
the apparent non-linear sequence of interference terms in the cohered 
system (such as the secondary beams or the sphere of diode light emitters) 
and their timing.   A second difficulty is in the memory allocation; 
one of the beams or a backup chip could be used for memory allocation 
in the role of a cache or larger memory designation with longer lifetime.

The beam configurations in the laser computing appear to generate 
computing with flops per second with extremely large numbers of digits.  
The architecture is an alternative to the conventional definition of a 
quantum computer.  Also the core of the computing doesn't involve 
nano-circuitry.

There exists much coherence stability protection not mentioned in this 
article, Jan 2006.

\newpage

\vskip 1 in.


\begin{thebibliography}{99}

\bibitem{QuantumComputing1}
Richard P. Feynman, ``Quantum Mechanical Computers,'' 
Plenary talk given at IQEC-CLEO Meeting, Anaheim, CA, Jun 19, 1984. 
In *Brown, L.M. (ed.): Selected papers of Richard Feynman* 968-992. 

\end{thebibliography}
\end{document}